\documentclass[aps,prb,twocolumn,groupedaddress,showpacs]{revtex4}
\usepackage{graphicx}
\usepackage{amssymb}

\begin{document}

\preprint{}

\title{ K-edge XANES of substitutional and interstitial Mn atoms
in (Ga,Mn)As}
\author{N.\ A.\ Goncharuk}
\affiliation{Institute of Physics, Academy of Sciences of the Czech
Republic,\nolinebreak[5] v.v.i.,\\ 
Cukrovarnick\'{a} 10, 162 53 Prague 6, Czech Republic}
\author{J.\ Ku\v{c}era}
\affiliation{Institute of Physics, Academy of Sciences of the Czech
Republic,\nolinebreak[5] v.v.i.,\\ 
Cukrovarnick\'{a} 10, 162 53 Prague 6, Czech Republic}
\author{K.\ Olejn\'{i}k}
\affiliation{Institute of Physics, Academy of Sciences of the Czech
Republic,\nolinebreak[5] v.v.i.,\\ 
Cukrovarnick\'{a} 10, 162 53 Prague 6, Czech Republic}
\author{V.\ Nov\'{a}k}
\affiliation{Institute of Physics, Academy of Sciences of the Czech
Republic,\nolinebreak[5] v.v.i.,\\ 
Cukrovarnick\'{a} 10, 162 53 Prague 6, Czech Republic}
\author{L.\ Smr\v{c}ka}
\affiliation{Institute of Physics, Academy of Sciences of the Czech
Republic,\nolinebreak[5] v.v.i.,\\ 
Cukrovarnick\'{a} 10, 162 53 Prague 6, Czech Republic}
\author{Z.\ Mat\v{e}j}
\affiliation{Charles University, Faculty of Mathematics and Physics,
  Department of Condensed Matter Physics,\\ 
Ke Karlovu 5, 121 16 Prague 2, Czech Republic}
\author{L.\ Nichtov\'{a}}
\affiliation{Charles University, Faculty of Mathematics and Physics,
  Department of Condensed Matter Physics,\\ 
Ke Karlovu 5, 121 16 Prague 2, Czech Republic}
\author{V.\ Hol\'{y}}
\affiliation{Charles University, Faculty of Mathematics and Physics,
  Department of Condensed Matter Physics,\\ 
Ke Karlovu 5, 121 16 Prague 2, Czech Republic}




\date{\today}

\begin{abstract}
This work reports theoretical and experimental study of the X-ray
absorption near-edge structure (XANES) at the Mn K-edge in (Ga,Mn)As
diluted magnetic semiconductors. The spectra have been calculated from
the first-principles using FLAPW including the core-hole effect, a
special attention has been paid to consequences of coexistence of Mn
impurities in substitutional and tetrahedral interstitial
positions. We have performed quantitative component analysis of
experimental spectra collected on the (Ga,Mn)As samples before/after
annealing and etching, with the aim to determine the proportion of Mn
impurity configurations. Comparison of the experimental data with
theoretical computations indicates that even after annealing and
etching some Mn atoms still reside in interstitial sites, although the
concentration of interstitial defects has been reduced by annealing.
\end{abstract}


\maketitle


\section{Introduction\label{Intro}}
Recently, the Mn-doped GaAs system has received considerable attention
in view of its potential use in spintronic technology, as it combines
both semiconducting/semimetallic and ferromagnetic properties in one
physical system \cite{Ohno1,Ohno2}.  The positions of Mn dopants play
a decisive role in determining the magnetic properties of (Ga,Mn)As.
There are three of them with a comparable energy \cite{Masek}.  The
substitutional Mn atoms, Mn$_{Ga}^{sub}$, occupying Ga sites, act as
hole-producing acceptors which contribute to ferromagnetism. The Mn
atoms in tetrahedral interstitial positions, surrounded by either As
or Ga atoms, Mn$_{As}^{int}$ and Mn$_{Ga}^{int}$, are
electron-producing double donors which hinder ferromagnetic
states. They partly compensate the Mn acceptors in the substitutional
positions and reduce the number of holes that mediate ferromagnetism
\cite{Jungwirth_review,Sinova_review}. Post-growth annealing of
(Ga,Mn)As samples at low temperatures, close to the growth
temperature, is known to improve magnetic properties of the (Ga,Mn)As
system
\cite{Yu,Edmonds1,Hashimoto,Kuryliszyn,Chiba,Stone,Limmer,Edmonds2}.
The concentration of Mn interstitials is reduced via out-diffusion
through the hexagonal interstitial sites towards a surface where they
oxidize. A Mn-rich oxide layer formed on the surface can be removed by
etching \cite{Olejnik}.

The X-ray absorption spectroscopy is the traditional tool which is
used to characterize the local environment of impurities
\cite{book-X1,book-X2}.  Not long ago a few works have been published
reporting on computations of XANES at the Mn K-edge in ternary
(Ga,Mn)As alloys \cite{Bacewicz,Titov1,Titov2,Acapito,Wolska} and in
related diluted magnetic semiconductors
\cite{Biquard,Sonoda,Sancho-Juan,Wei}.  However, most of these XANES
studies concern mainly substitutional Mn defects. In the present paper
a special attention is paid to interpretation of X-ray absorption
spectra measured on Mn-doped GaAs materials in which different types
of impurities coexist.

Let us note that recently detailed investigations focused on the
configuration of substitutional and interstitial Mn impurities in
(Ga,Mn)As have been performed with cross-sectional scanning tunneling
microscopy (XSTM) \cite{Stropa} and electron paramagnetic resonance
(EPR)\cite{Weiers}.

Our study is based on the comparison of experimental XANES with
spectra calculated by the method of full potential linearized
augmented plane waves (FLAPW) \cite{wien2k}. An isolated defect is
represented by a supercell which should be as large as possible to
describe correctly the absorption process. Therefore, we simulate the
spectra from different types of defects by separated calculations,
considering always only one substitutional or interstitial atom in a
supercell.

If we consider an isolated Mn impurity, the position of the Fermi
energy on the energy scale depends on the defect type, i.e., it is
fixed by the acceptor level near the top of the valence band for the
substitutional atom and by the donor levels close to the bottom of the
conduction band for the interstitial atoms, respectively. The
situation is similar in the supercell scheme with a single defect in
each supercell. The concentration of defects is finite, the
acceptor/donor levels are broadened and merge with the
valence/conduction band, but the Fermi energy remains fixed near the
valence or conduction band, depending on the defect type.  This
complicates the quantitative analysis of experimental absorption
spectra on the basis of the linear combination of theoretical spectra
calculated separately for each impurity.

In the real crystal with a mixture of defects the position of the
Fermi energy is determined by the type of defects with the higher
concentration, i.e., in the ferromagnetic (Ga,Mn)As by concentration
of acceptors. We model this situation by replacing the Fermi energy of
supercells with donors by the Fermi energy of the supercell with an
acceptor. This shift of the Fermi energy must be respected in the
linear combination of spectra fitting of the experimental data.

\section{Experiment \label{Exp}}
The measurements have been carried out at the European Synchrotron
Radiation Facility in Grenoble, beamline BM29. 
The XANES spectra have been obtained by measuring the Mn K$_{\alpha}$
fluorescence as a function of the energy of incident photons around
the energy of the Mn K absorption threshold, $E_{\circ} = 6539$~eV.

The measurements have been performed in the grazing-incidence mode,
in which the penetration depth of the primary X-ray beam was
sensitively tuned by changing the incidence angle $\alpha$ of the
primary radiation in the range from 0.1$^{\circ}$ to 0.45$^{\circ}$,
i.e., around the critical angle $\alpha_{c}\approx 0.38^{\circ}$ of
total external reflection. Increasing the incidence angle in this
range, the penetration depth grows from few nanometers to several
microns. 

The fluorescence method of measurement was employed. The measured
fluorescence signal comes from the transition of $2p$ L-shell core-level
electrons to $1s$ K-shell core holes created by primary X-ray wave. The
energy of the corresponding K$_{\alpha}$ fluorescence line is
characteristic of the Mn atom and its intensity is proportional to the
absorption coefficient (the XANES signal).  Spectra were measured at
room temperature.

The sample grown by molecular beam epitaxy (MBE) with $\sim 10.5-11~\%$
of Mn content was used in our experiments. After annealing, the
content of Mn was reduced to $\sim 7.5-8~\%$. The annealed sample was
etched in $35~\%$ HCl.  Manganese K-edge XANES spectra were taken in
fluorescence from the 42~nm thick sample of the as-grown, annealed and
etched-after-annialing (Ga,Mn)As.

\begin{figure}[ht]
\begin{center}
\includegraphics[width=\linewidth]{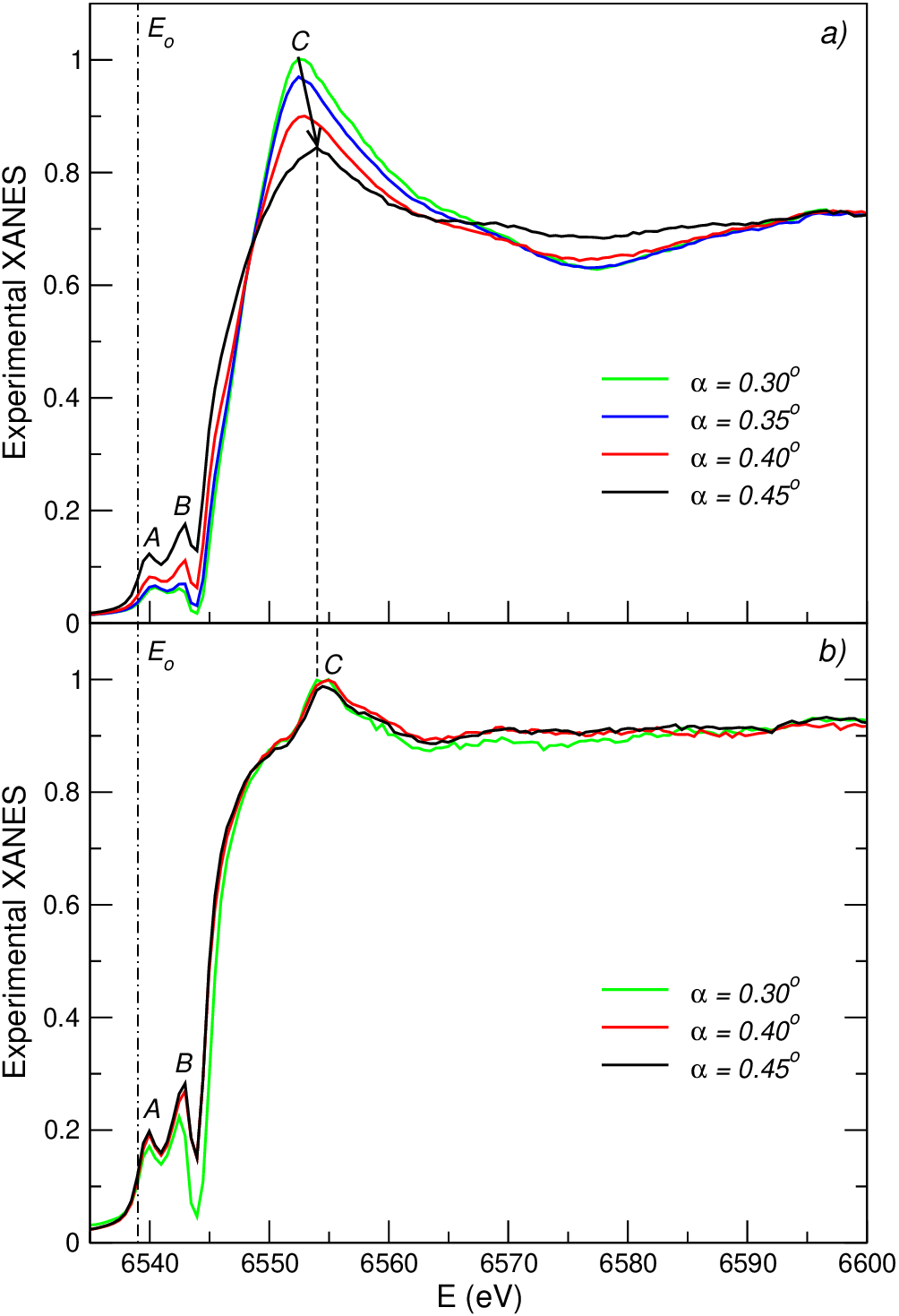}
\caption{\label{experiment} 
(Color online) Experimental K-edge X-ray absorption spectra
  of Mn in (Ga,Mn)As measured at different incidence angles $\alpha$
  of the primary X-ray beam. {\it a)} Spectra taken on the annealed
  sample with the Mn-rich surface oxide layer. {\it b)} Spectra taken
  on the annealed-and-etched sample without the Mn-rich surface oxide
  layer. All the spectra are normalized with respect to their
  high-energy ends. The different structures labeled A (pre-edge), B
  (pre-edge) and C (main absorption peak) are discussed in the text.}
\end{center}
\end{figure}

Two sets of experimental Mn K-edge XANES spectra, measured on the annealed and
annealled-and-etched (Ga,Mn)As samples, are presented in
Fig.~\ref{experiment} for various incidence angles $\alpha$. 
The spectra can be separated into two parts: the pre-edge and main
absorption. The pre-edge region, peaks A and B at around $E\sim
6540$~eV, lies below the steeply rising absorption edge, which
culminates in the most intense post-edge peak C in the vicinity of
$E\sim 6552\div 6554$~eV, as illustrated in Fig.~\ref{experiment}.  All
experimental spectra, including not presented here as-grown ones, exhibit
the double structure (peaks A and B) separated by $\sim 2.8$~eV in the
pre-edge absorption region.

In the annealed sample measured at lower $\alpha$, the absorption edge
grows almost linearly [Fig.~\ref{experiment}{\it a)}]. With $\alpha$
increased, the spectra in this region acquire a convex shape and the
position of the peak C moves to higher energies.  The spectra taken at
high $\alpha$ remind those measured on the annealed-and-etched sample
[Fig.~\ref{experiment}{\it b)}]. This spectrum modification must be
due to removing the oxide layer from the sample surface by etching.

\section{Computational model}
\label{CM}

The theoretical Mn K-edge XANES spectra and electronic band structures
of (Ga,Mn)As were simulated using the full potential linearized
augmented plane wave plus local orbitals (FLAPW + lo)
method\cite{xanes-T1,xanes-T2} implemented by Blaha et al.  in the
WIEN2k package \cite{wien2k}.  A generalized gradient approximation
(GGA) \cite{Perdew} was employed for the exchange-correlation
functional.

Basis wave functions were expanded in combinations of spherical
harmonics inside non-overlapping muffin-tin (MT) spheres surrounding
the atomic sites and in Fourier series in the interstitial region,
with a cutoff of R$_{\textrm{MT}}$K$_{\textrm{max}}$ fixed at 7.0,
where R$_{\textrm{MT}}$ denotes the smallest MT sphere radius and
K$_{\textrm{max}}$ is the magnitude of the largest $k$-vector in the
plane wave expansion. The MT radii were assumed to be 1.29, 1.22 and
1.14~\accent23 A for Mn, Ga and As, respectively.

Brillouin zone integrations were performed using a tetragonal
$k$-point mesh of Monkhorst-Pack type \cite{Monkhorst}.  Convergence
of self-consistency was obtained using $5\times 5 \times 5$ $k$-point
sampling mesh in the reciprocal space of each supercell that
corresponds to 10 $k$-points in the irreducible wedge of the Brillouin
zone.

The spin-polarized calculations were performed within a 64-atom
supercell based on the zinc-blende GaAs cubic cell with the
experimental lattice constant $a=5.65$~\accent23 A, doped with
3.125/3.03 atomic percent of Mn impurities in the
substitutional/interstitial site.  We always assumed a single Mn atom
inside each supercell and used unrelaxed positions of the nearest
neighbors around Mn.  In the substitutional position the Mn atom is
surrounded by 4 As with the unrelaxed Mn-As bond distance
2.445~\accent23 A, the second neighbors are 12 Ga on the distance
3.994~\accent23 A. When Mn is in one of interstitial positions, its
first neighbors are either 4 As or 4 Ga, and the second neighbors are
either 6 Ga or 6 As, respectively, with bond lengths equal to
2.445~\accent23 A and 2.822~\accent23 A.  Thus, we expect that the
difference between the energy structure in two configurations,
Mn$_{Ga}^{sub}$ and Mn$_{As}^{int}$, in which Mn is surrounded by 4 As
atoms is determined mainly by the second nearest neighbors.

\begin{figure}[htb]
\begin{center}
\includegraphics[width=0.9\linewidth,angle=0]{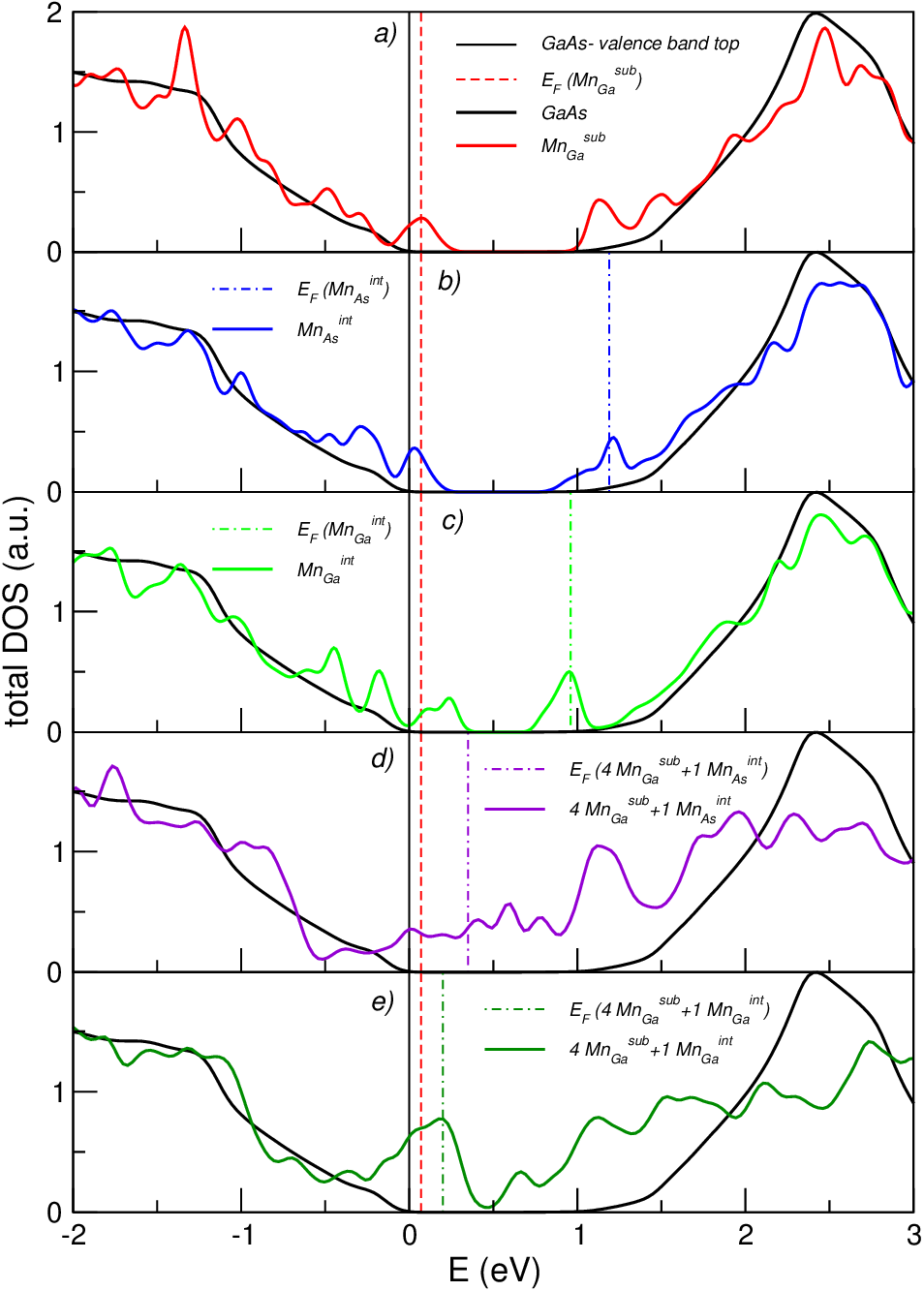}
\caption{\label{DOS-GaAs}
The total DOS of GaAs (black curve) together with
  total DOSs of (Ga,Mn)As with embedded 
{\it a)} single substitutional Mn atom, 
{\it b)} single interstitial Mn atom inside the As tetrahedron,  
{\it c)} single interstitial Mn atom inside the Ga tetrahedron, 
{\it d)} cluster of four substitutional Mn atoms and one
interstitial Mn atom inside the As tetrahedron, and
{\it e)} cluster from four substitutional Mn atoms and one
interstitial Mn atom inside the Ga tetrahedron. 
The full horizontal line at $E=0$~eV denotes the top of the valence
band. The Fermi energies of particular supercells 
are indicated by dashed vertical lines. }
\end{center}
\end{figure}
The total density of states (DOS) calculated for three supercells are
shown in the top three panels of Fig.~\ref{DOS-GaAs} together with the
DOS of GaAs and positions of the Fermi levels, which reflect the
presence of either acceptor or donor impurities. As mentioned in
Sec.~\ref{Intro}, the Fermi level in ferromagnetic (Ga,Mn)As is
determined by prevalence of acceptors, and situated in the band
composed of the merged valence and broadened acceptor impurity band.
Consequently, we will consider the states above
the Mn$_{Ga}^{sub}$ Fermi level as empty also for DOS calculated for
Mn$_{As}^{int}$ and Mn$_{Ga}^{int}$ when interpreting XANES spectra.
  
We support this approach by calculation of the Fermi level positions
in unrealistic 64-atom supercells with incorporated 4
Mn$_{Ga}^{sub}$ and 1 Mn$_{As/Ga}^{int}$. The results are shown in two
bottom panels of Fig.~\ref{DOS-GaAs}. As expected, the resulting Fermi
energies are shifted closer to the value obtained for the supercell
with a single substitutional impurity.

In the single-electron approximation, which is used in our model, the
absorption spectrum is given by a transition between the initial
ground state with fully occupied core-levels and the final state with
one core electron removed.  The core-hole effect is known to be quite
important for reproducing experimental spectra by theoretical
calculations \cite{Yamamoto,Nakashima}. Therefore, we repeated our
selfconsistent calculations also for excited states with one core
electron removed from the Mn $1s$ level and an additional valence
electron placed close to the bottom of the conduction band to fill the
lowest unoccupied $4p$ orbital.

The energy of the absorbed radiation is determined as a difference
between the calculated total energies of the initial and final
states. Note, that the influence of Mn $1s$ core-level shifts due to
adjacent As or Ga atoms is included automatically in the self
consistent calculation.  For the supercells with interstitial Mn atoms
this energy must be further corrected by the Fermi level shifts, as
mentioned above.

Due to the very small radius of Mn $1s$ K-shell core hole the
absorption spectrum is given as a product of the transition
probability, calculated from $1s$ - $4p$ dipole matrix elements, which
is a smooth function of energy, and the local partial DOS of
$4p$-states above the Fermi energy in the MT sphere surrounding the Mn
atom, obtained from the final state calculation.
\section{Data analysis and results}
\label{Results}
\subsection{First-principles XANES simulations}
\label{FPS}
\begin{figure}[b]
\begin{center}
\includegraphics[width=\linewidth,angle=0]{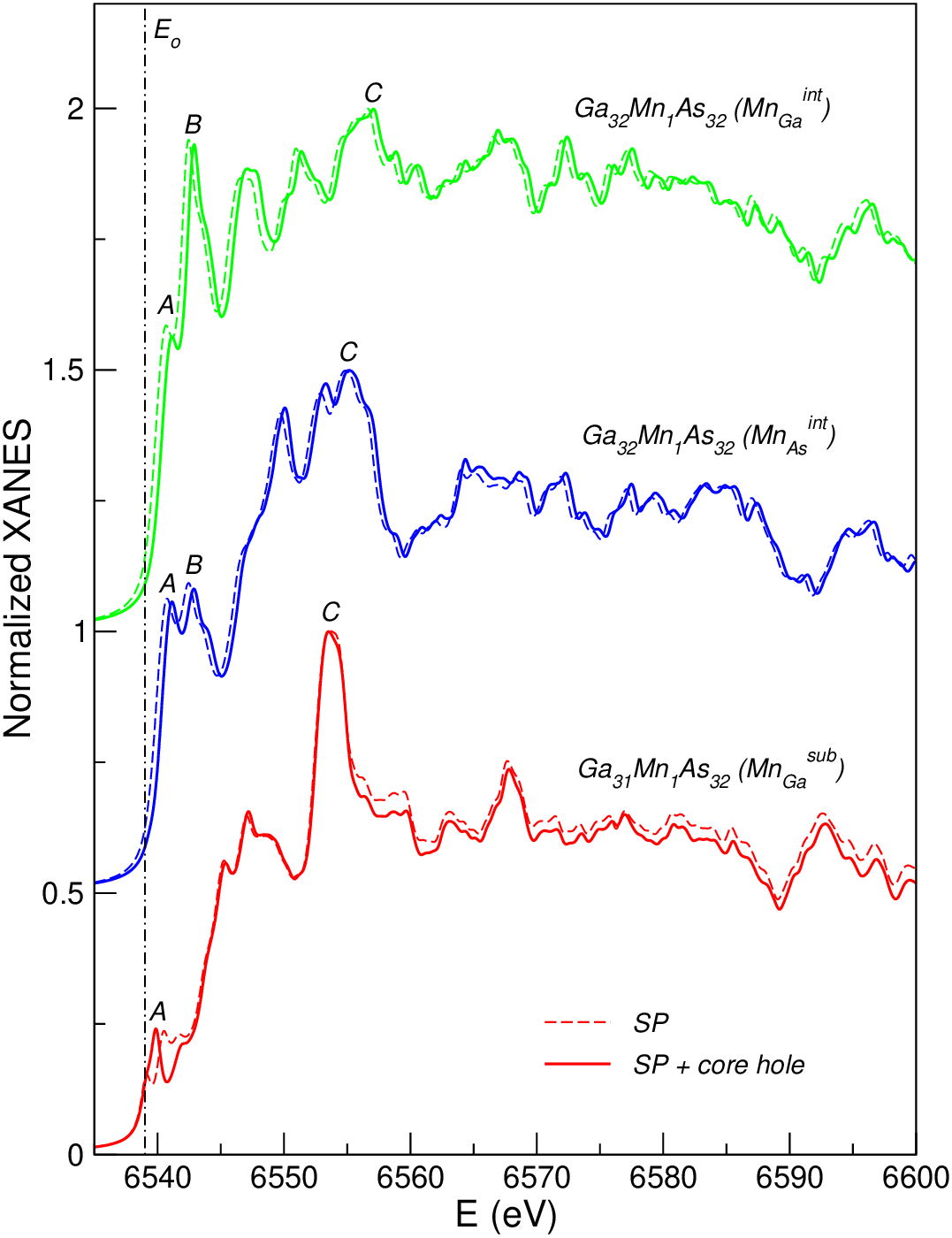}
\caption{\label{xanes}
Normalized ab-initio K-edge XANES spectra of Mn in the 64-atom (Ga,Mn)As
supercell simulated for three diferent Mn sites in the substitutional 
(Mn$_{Ga}^{sub}$) and two tetrahedral interstition positions 
(Mn$_{As}^{int}$ and Mn$_{Ga}^{int}$). 
Solid/dashed curves are spin-polarized spectra calculated with/without 
the core-hole effect.
}
\end{center}
\end{figure}
The simulated XANES Mn K-edge spectra for three types of isolated Mn
defects are reported in Fig.~\ref{xanes}. All curves were normalized
and convoluted with a Lorentz function modeling the experimentally
induced broadening. A parameter of a full width at half maximum (FWHM)
$\gamma = 1$~eV was chosen for a fair comparison between experimental
and calculated spectra. The use of the broadening with larger $\gamma$
would mask the fine details of theoretical curves in the pre-edge
region.

The transition energy was obtained by a difference in total electronic
energies between the ground and core-hole states.  The theoretical
transition energy deviates from the experimental one (threshold,
$E_{\circ}= 6539$~eV) by $-22$~eV $\left(|\Delta E/E| =
0.336\%\right)$, which is reflected in Fig.~\ref{xanes} by the
corresponding shift of the energy scale.

Similarly to the experimental XANES spectra, the pre-edge and
main absorption parts can be distinguished. An intense
double structure (peaks A and B) appears in
the pre-edge region for both types of interstitial defects,
Mn$_{As}^{int}$ and Mn$_{Ga}^{int}$, in contrast to a weak single peak
A for a substitutional Mn impurity, Mn$_{Ga}^{sub}$.

Arrangement of neighboring atoms around the absorbing Mn atom strongly
affects the pre-edge structure.
The intensity of pre-edge peaks is much larger for interstitials
compared to that for the substitutional model spectrum. In accordance 
with Wong \cite{Wong}, we attribute this difference to the different
distance of the second neighbor shell ligands from the X-ray absorbing
Mn centers. As already mentioned in Sec.~\ref{CM}, all considered Mn
impurities are located inside tetrahedrons composed of As/Ga atoms
with the same bond length of the first neighbor shell
ligands. However, the bond distance of the second nearest neighbor
ligands of interstitials is by 1.172~\accent23 A less than that of
Mn$_{Ga}^{sub}$.  In the Wong's terminology, the smaller the
``molecular cage'', the higher the intensity of the pre-edge
absorption.

As for the core-hole influence on the shape of the calculated spectra,
our results imply that it is significant only in the pre-edge region
of the substitutional Mn defect and is not important in the main part
of absorption curves.  With the core-hole effect introduced, the
spectral structure of both type of interstitials was only shifted by
$-0.4$~eV but its shape remained the same.  Let us now discuss the the
pre-edge electronic structure in more detail.
\begin{figure}[t]
\begin{center}
\includegraphics[width=\linewidth]{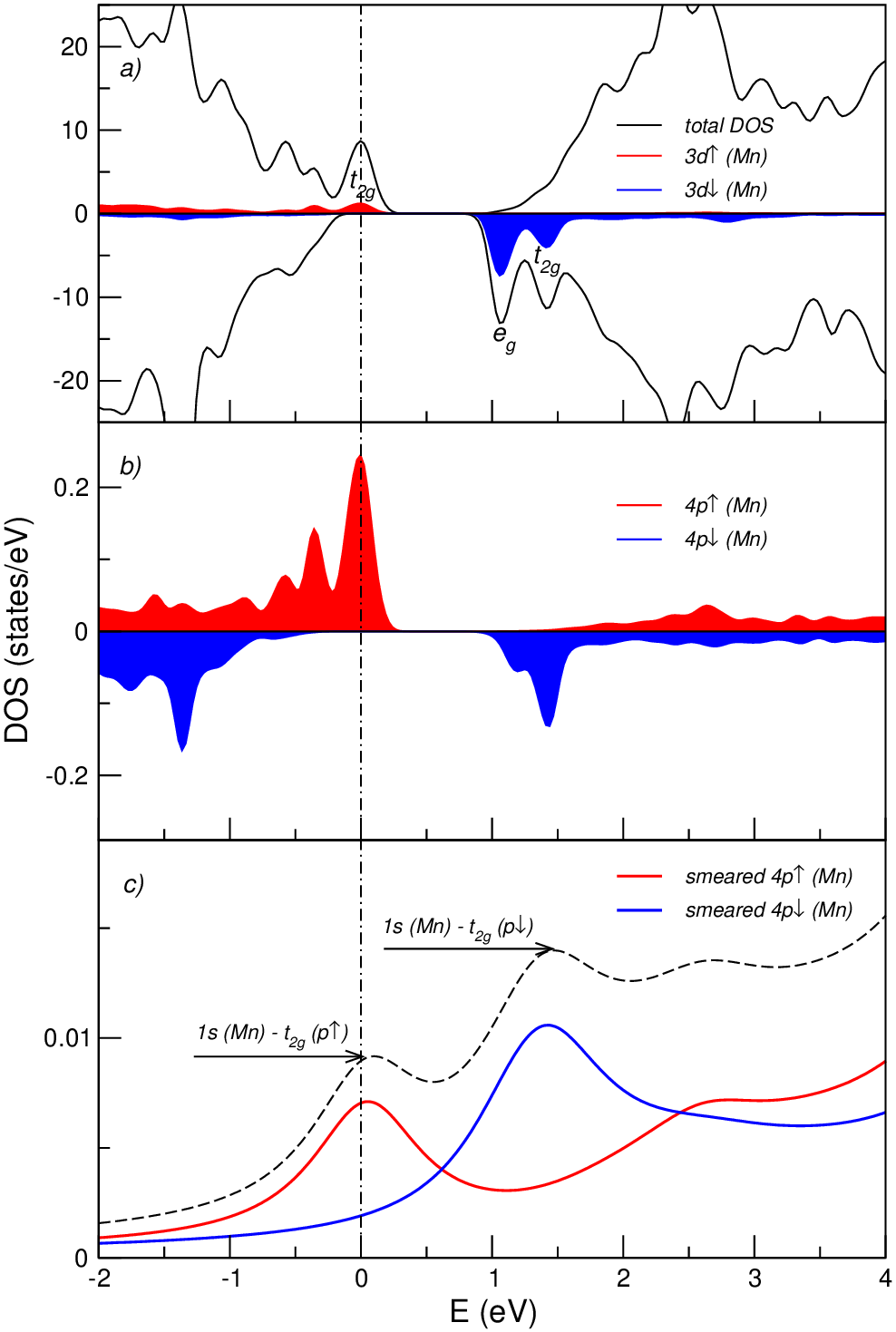}
\caption{
(Color online) The total and partial DOSs in the 
Ga$_{31}$Mn$_{1}$As$_{32}$ supercell with a Mn atom in the substitutional
site calculated without the core-hole effect. 
{\it a)} The total DOSs (solid curves) and densities of Mn
$3d$-states (filled regions). 
{\it b)} The density of $4p$-states of Mn.
{\it c)} The smeared density of $4p$-states of Mn 
(the X-ray absorption spectrum divided by the corresponding matrix element).   
The majority/minority spin DOSs are shown in the upper/lower
part of figures. The vertical dash-dot line denotes the Fermi level. 
The electron transitions in the pre-edge part of the
absorption spectrum are depicted by arrows.} 
\label{DOS-Mn-sub}
\end{center}
\end{figure}
\begin{figure}[t]
\begin{center}
\includegraphics[width=\linewidth]{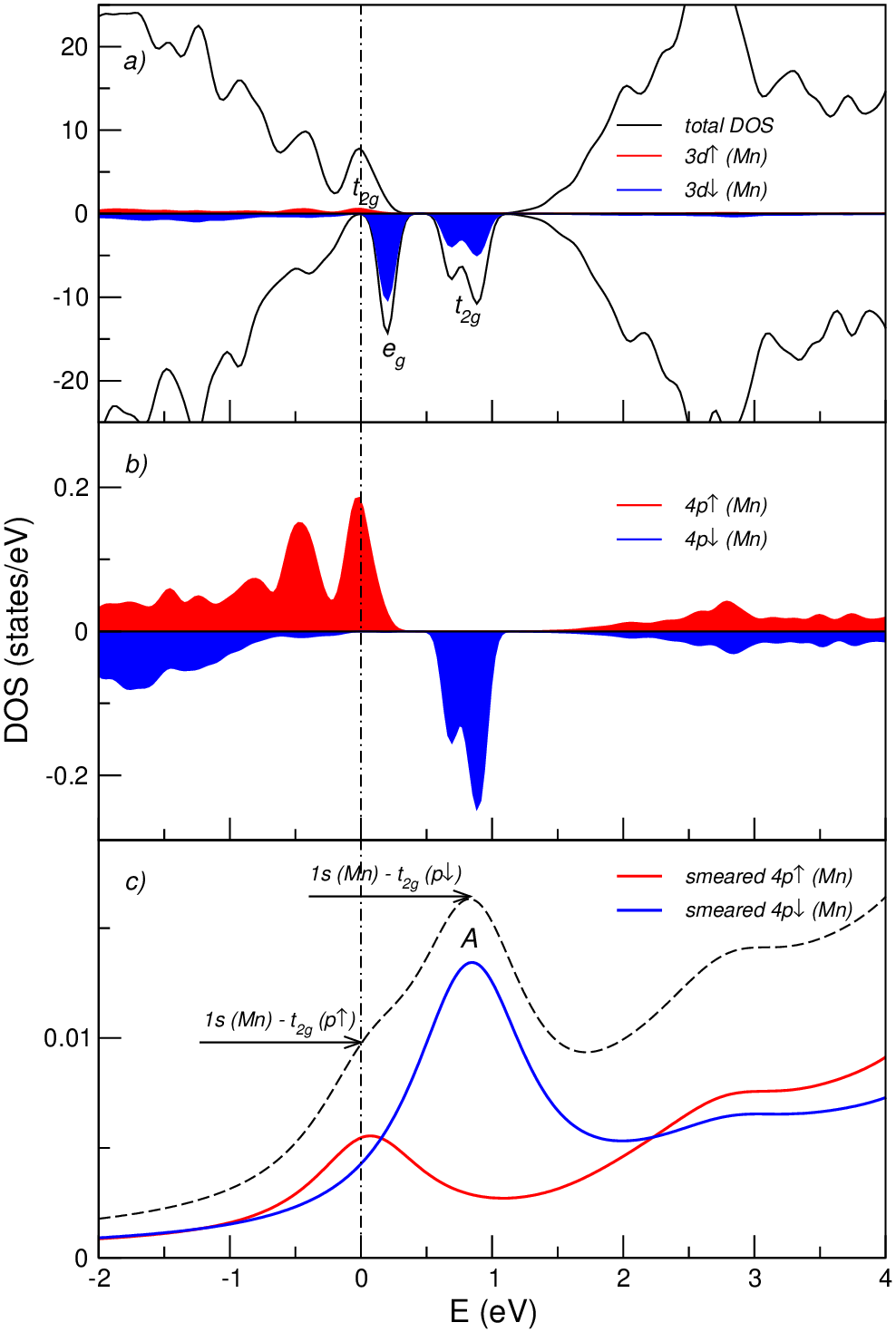}
\caption{
(Color online) The same as in Fig.~\ref{DOS-Mn-sub} calculated with
  the core-hole effect.} 
\label{DOS-Mn-sub-HOLE}
\end{center}
\end{figure}
\begin{figure}[t]
\begin{center}
\includegraphics[width=\linewidth]{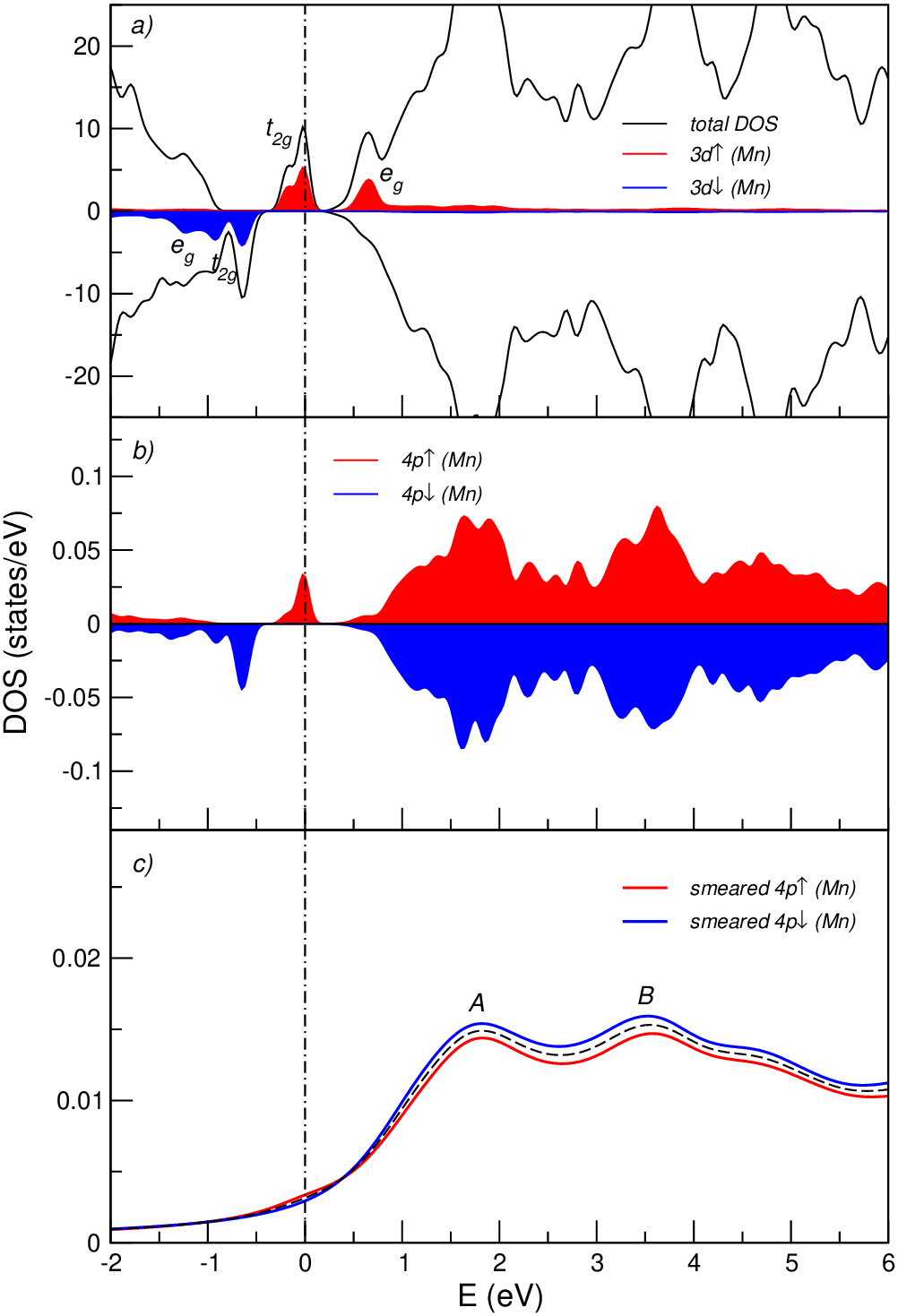}
\caption{\label{DOS-Mn-intAs-HOLE} 
(Color online) The same as in Fig.~\ref{DOS-Mn-sub} for the 
Ga$_{32}$Mn$_{1}$As$_{32}$ supercell with the interstitial Mn atom 
inside the As tetrahedron calculated with the core-hole effect.}
\end{center}
\end{figure}
\begin{figure}[t]
\begin{center}
\includegraphics[width=\linewidth]{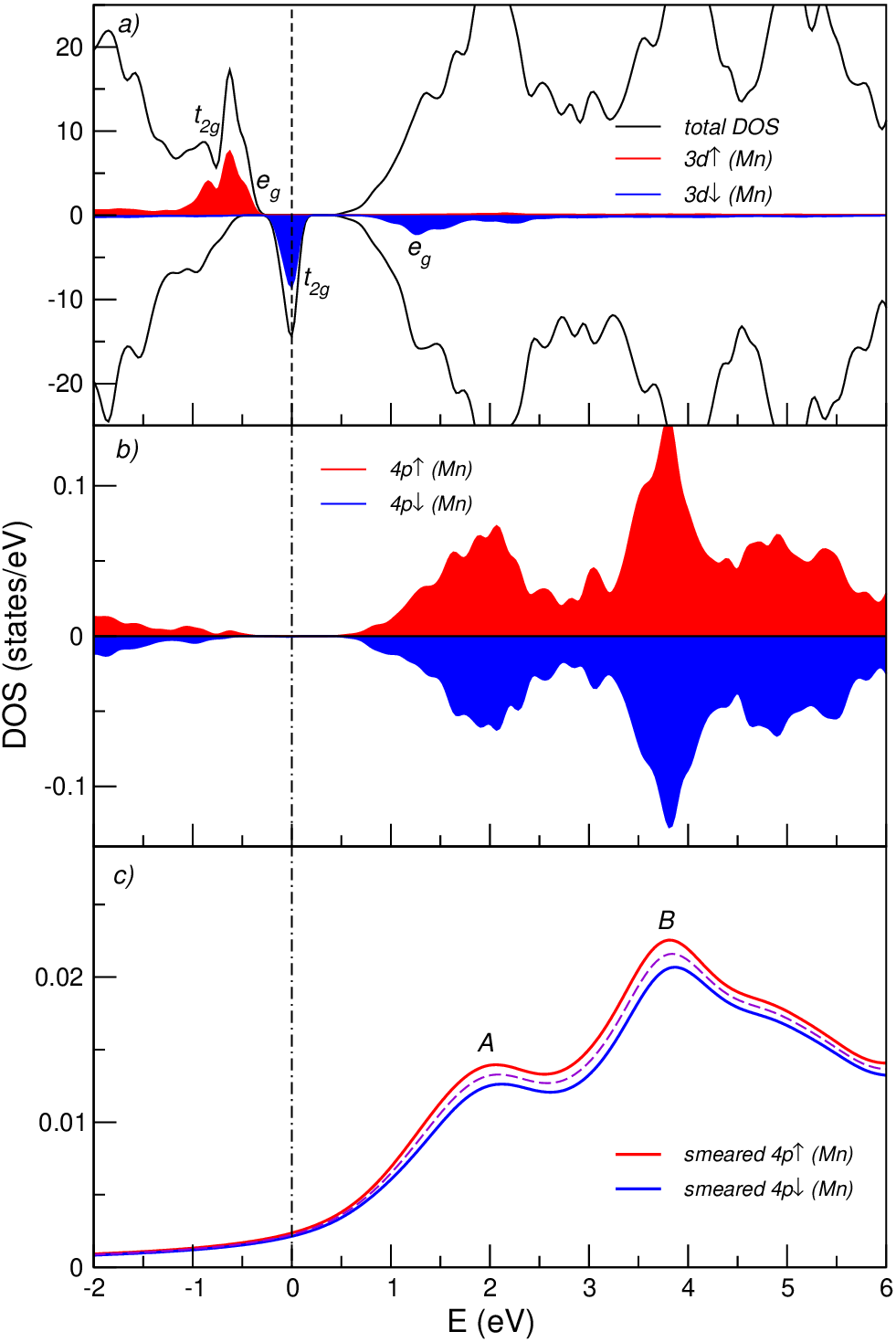}
\caption{\label{DOS-Mn-intGa-HOLE} 
(Color online) The same as in Fig.~\ref{DOS-Mn-sub} for the   
Ga$_{32}$Mn$_{1}$As$_{32}$ supercell with the interstitial Mn atom 
inside the Ga tetrahedron with the core-hole effect. }
\end{center}
\end{figure}
 
The band structure of (Ga,Mn)As with Mn$_{Ga}^{sub}$ calculated
without/with taking into account the core-hole effect is shown in
Figs.~\ref{DOS-Mn-sub},\ref{DOS-Mn-sub-HOLE}. When Mn atom is in the
substitutional position it is surrounded by four As atoms located on
the vertices of a tetrahedron.  The tetrahedral crystal field of As
ions splits $3d$-states of Mn into $e_{g}$- and $t_{2g}$-levels with
$e_{g}$ below $t_{2g}$ \cite{Zunger1,Sato,Zunger2}. Exchange
interactions further split these states into spin-up ($\uparrow$) and
spin-down ($\downarrow$) states. Due to the tetrahedral arrangement
the $t_{2g}$-states of the Mn atom hybridize with Mn $4p$-states. The
states resulting from hybridization of the deeper
$t_{2g}^{\uparrow}$-states have mainly Mn $d$ character, while those
resulting from higher $t_{2g}^{\downarrow}$-states have dominantly As
$p$ character.  In contrast, $e_{g}$-levels of Mn have no states
available for significant coupling since the GaAs host does not have
$e_{g}$-states localized in this energy range. It is seen in
Fig.~\ref{DOS-Mn-sub} that the density of Mn $p$-states is larger at
the energy of $d_{t_{2g}}$-states in comparison with that at the
energy of $d_{e_{g}}$-states.

It is obvious from the discussion above that the pre-edge fine
structure of Mn K-edge XANES with Mn$_{Ga}^{sub}$ originates from the
transfer of $1s$ electrons to valence $3d$ states mediated by $1s$--$4p$
dipole transition and $4p$--$3d$ hybridization.  We observed two weak
absorption lines corresponding to $1s$--$t_{2g}^{\uparrow}$ and
$1s$--$t_{2g}^{\downarrow}$ electron transfers in the absorption
spectrum calculated without the core-hole correction [see
Fig.~\ref{DOS-Mn-sub} {\it c)}].  The first transfer is only partly
available as the Fermi level touches the $t_{2g}^{\uparrow}$-band near
its border, leaving a little amount of empty
$t_{2g}^{\uparrow}$-states.  After introduction of the core-hole
effect, two aforementioned lines join into one line of higher
intensity [see Fig.~\ref{DOS-Mn-sub-HOLE} {\it c)}], and the transfer
$1s$--$t_{2g}^{\downarrow}$ becames dominant.  In this case the Mn
$t_{2g}^{\uparrow}$-states are fully filled with electrons and the
transfer $1s$--$t_{2g}^{\uparrow}$ becomes impossible.
   
The band structures of (Ga,Mn)As calculated with incorporated
interstitials Mn$_{As}^{int}$ and Mn$_{Ga}^{int}$, and taking into
account the core-hole effect, are shown in
Figs.~\ref{DOS-Mn-intAs-HOLE},\ref{DOS-Mn-intGa-HOLE}. For both
interstitials, we observe two intense pre-edge absorption lines A and
B, which are originated from the $1s$--$4p$ transitions. This is in
contrast to the pre-edge structure of Mn$_{Ga}^{sub}$. Both peaks are
mainly of dipolar origin.  The Fermi level separates occupied and
unoccupied states. In the case of interstitial impurities it is fixed
in the band composed of the merged conduction and broadened donor
impurity band. The Fermi level falls in the $t_{2g}^{\uparrow}$-band
for Mn$_{As}^{int}$ and $t_{2g}^{\downarrow}$-band for
Mn$_{Ga}^{int}$. However, these two bands do not contribute to the
pre-edge absorption structure due to low density of Mn $3d$-states
near the Fermi level. In contrast to the case with the substitutional
Mn, the $t_{2g}^{\uparrow}$/$t_{2g}^{\downarrow}$-states near the
Fermi level do not participate in $4p$--$3d$ hybridization. Thus, both
peaks are mainly of dipolar origin, and their shape replicates the DOS
of Mn $4p$-states. 

\subsection{Experiment versus theory}
As already demonstrated in Sec.~\ref{Exp}, 
all experimental spectra (as-grown, annealed, and
annealed-and-etched) have got the double structure in the
pre-edge absorption. 
In Sec.~\ref{FPS} we showed that all examined Mn
impurities can contribute to the first pre-edge peak A. Only Mn
interstitials are responsible for the intensity growth of the second
pre-edge peak B. It implies that at least small proportion of interstitial Mn
defects is present in the compound.
\begin{figure}[t]
\begin{center}
\includegraphics[width=\linewidth,angle=0]{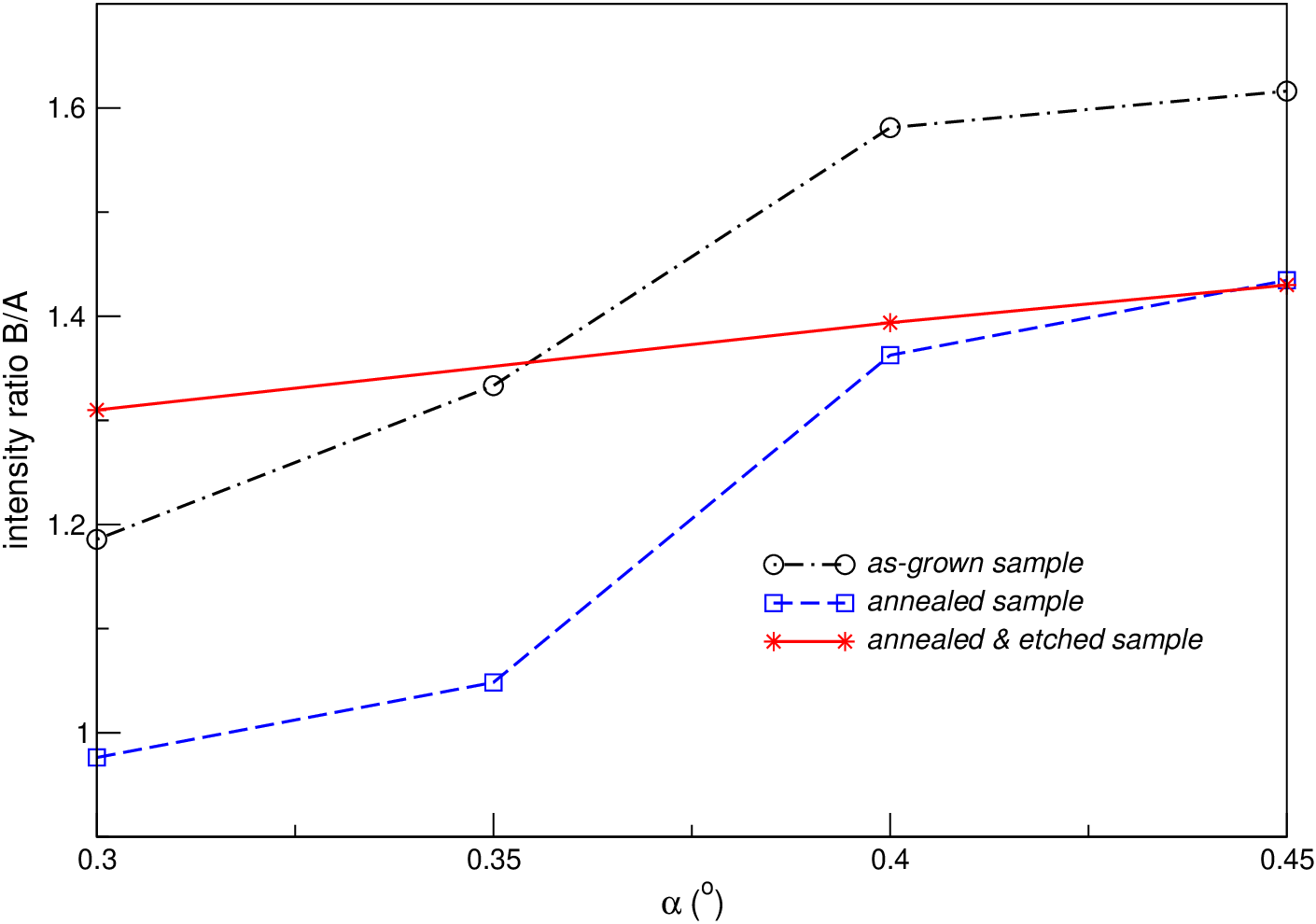}
\caption{\label{ratio_B_A}
The ratio of experimental pre-edge peak intensities B
  and A, obtained from measurements on the as-grown, annealed and
  annealed-and-etched sample as a
  function of the incidence angle $\alpha$ of the primary X-ray radiation.}
\end{center}
\end{figure}
The intensity ratios of experimental peaks B and A differs for
different incidence angles of the X-ray radiation, $\alpha$. The
tendency of B/A variation deep inside the sample is
displayed in Fig.~\ref{ratio_B_A}. Close to the surface,
$\alpha=0.3^{\circ}$, the ratio B/A, i.e., the content of Mn atoms in
interstitial positions, is smaller in the as-grown and annealed
samples, compare to the etched one. It is believed that the as-grown and
annealed samples have a thick layer of Mn oxide on the surface, and
the penetration depth of the X-ray beam inside the sample is small
at small $\alpha$, as it is reduced by the width of the oxide layer. 
In the annealed-and-etched sample the ratio B/A grows slowly with
$\alpha$. Thus, the Mn defect distribution is almost depth independent. 
The content of Mn impurities becomes equal in both annealed and
etched samples at large $\alpha = 0.45^{\circ}$ deep inside the sample
where the oxid layer is of no importance.

\begin{figure}[b]
\begin{center}
\includegraphics[width=\linewidth,angle=0]{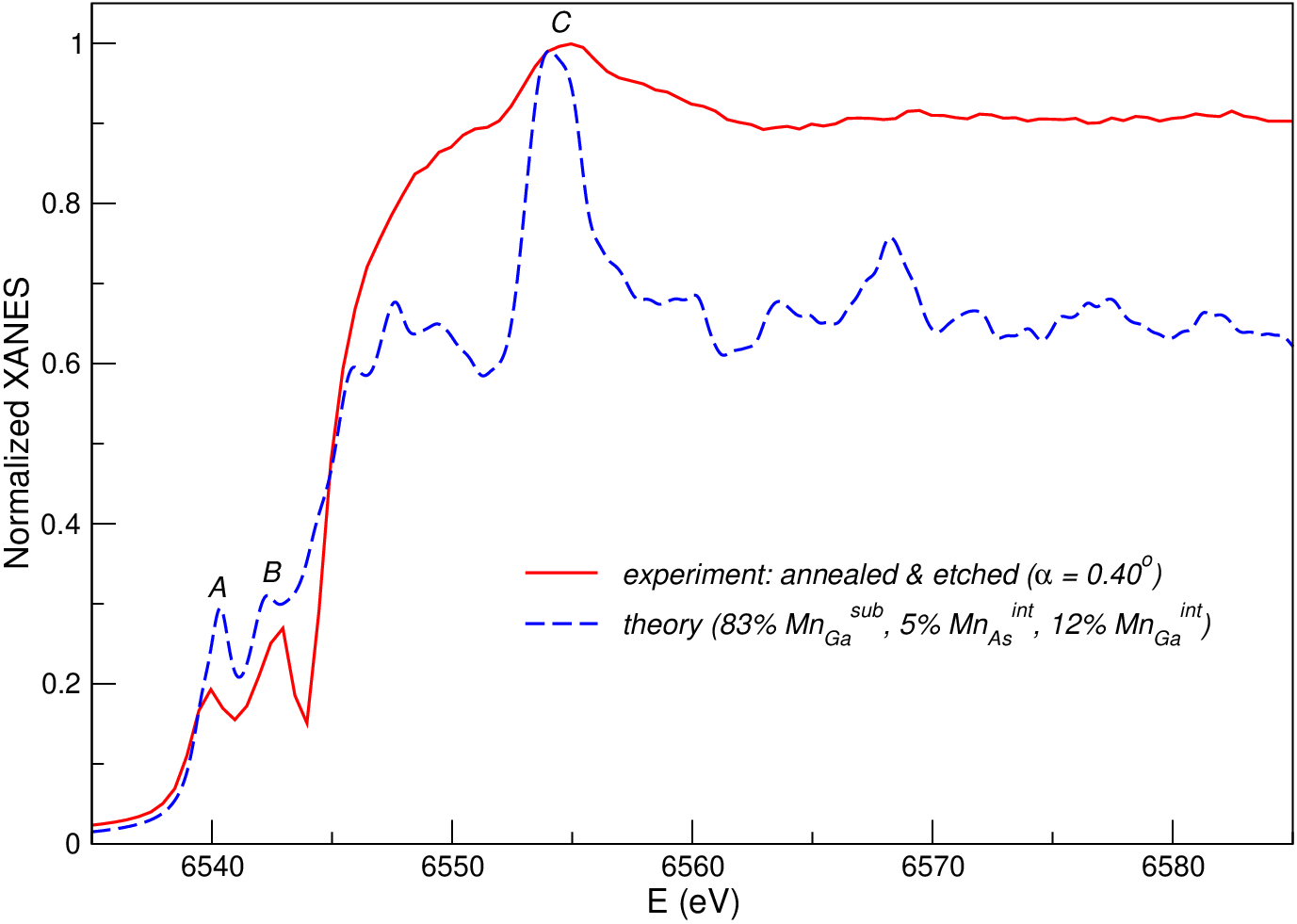}
\caption{\label{exp-etch}
Comparison of the normalized experimental and 
theoretical K-edge XANES spectra of Mn in (Ga,Mn)As. 
The experimental spectrum (solid curve)
was measured on the annealed-and-etched sample without the Mn-rich
surface oxide layer at $\alpha=0.40^{\circ}$. The
theoretical spectrum (dashed curve) 
is obtained by a linear combination of the model XANES
spectra calculated for Mn$_{Ga}^{sub}$, Mn$_{As}^{int}$ and
Mn$_{Ga}^{int}$ site defects.}
\end{center}
\end{figure}
\begin{figure}[t]
\begin{center}
\includegraphics[width=\linewidth,angle=0]{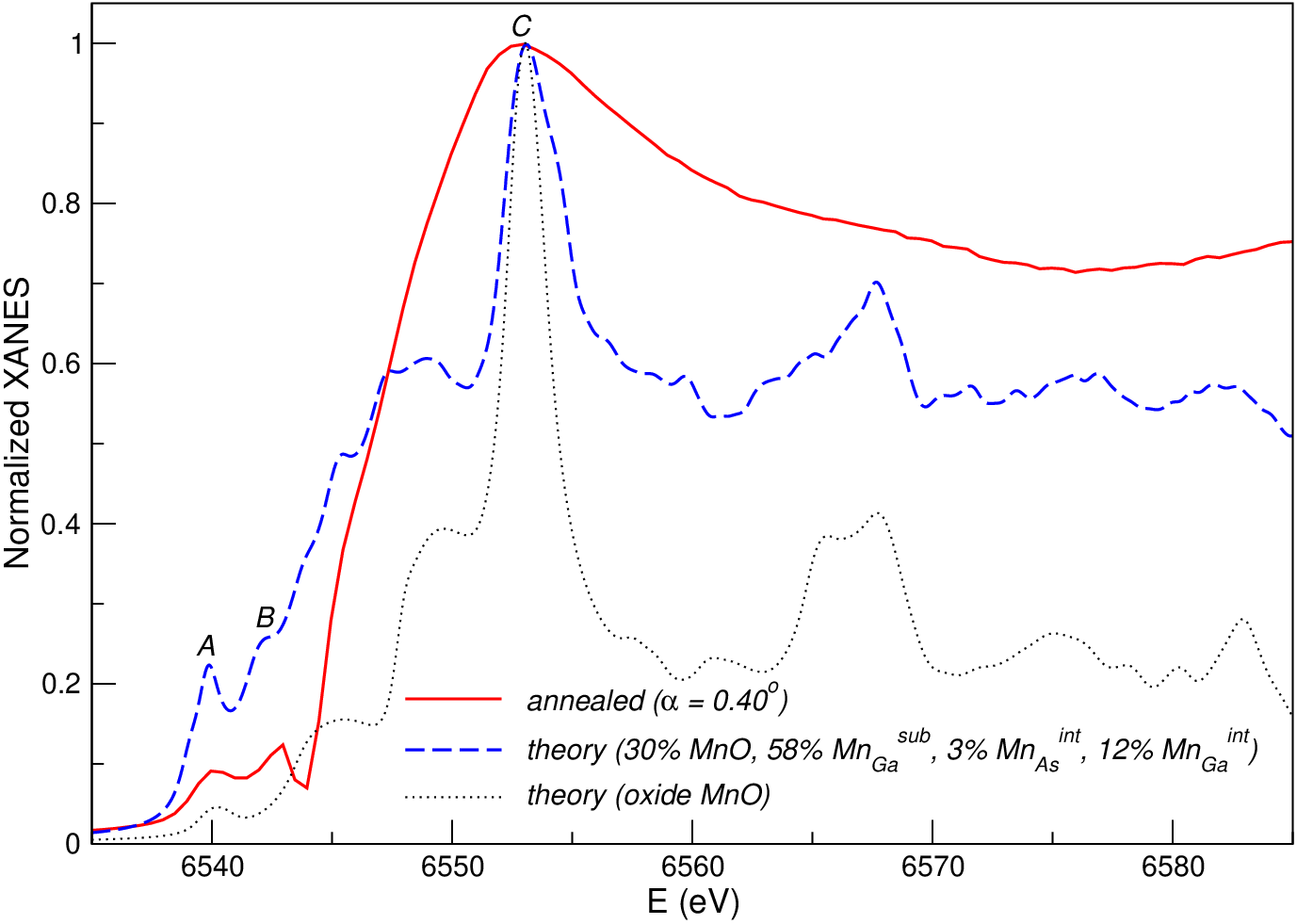}
\caption{\label{exp-anneal} 
Comparison of the normalized experimental and theoretical K-edge XANES
spectra of Mn in (Ga,Mn)As.
The experimental spectrum (solid curve)
was measured on the annealed sample with the Mn-rich
surface oxide layer at $\alpha=0.40^{\circ}$. 
The theoretical spectrum (dashed curve) is obtained by a 
linear combination of the model XANES spectra calculated 
for Mn$_{Ga}^{sub}$, Mn$_{As}^{int}$, Mn$_{Ga}^{int}$ site defects and
MnO oxide. The latter spectrum is shown (dotted curve).}
\end{center}
\end{figure}
The simulated data have been processed to explain the experimental
XANES spectra. Our aim was to guess a rough number of Mn
impurities that can reproduce the experimental spectra. Results are
present in Figs.~\ref{exp-etch},\ref{exp-anneal}.

Before performing a fit we fixed the Fermi energy of the supercell
with embedded substitutional Mn atom, $E_F(Mn_{Ga}^{sub}) = 4.609$~eV.
The absorption spectra obtained for Mn interstitials were shifted
relative to the spectrum calculated for the substitutional Mn by $\Delta
E_F = E_F(Mn_{Ga}^{sub})- E_F(Mn_{As/Ga}^{int})$, which is $-0.831$~eV
(for Mn$_{As}^{int}$) and $-0.578$~eV (for Mn$_{Ga}^{int}$),
correspondingly.

We found that the content of Mn$_{Ga}^{sub}$ prevails over the content
of both interstitial impurities, especially deep inside the sample
($\alpha \ge 0.40^{\circ}$). We have got 83$\%$ of Mn$_{Ga}^{sub}$ and
17$\%$ of interstitials. The intensity of the peak B can be changed by
the variation of the proportion of Mn$_{As}^{int}$ and
Mn$_{Ga}^{int}$. The more Mn$_{Ga}^{int}$, the higher the feature B. 
Analogously to the experimental curve, the theoretical fit has a convex 
shoulder above the absorption edge, which has been observed only on
the annealed-and-etched sample [Fig.~\ref{experiment} {\it b)}].     

Our fit is not exact which indicates shortages of the component
analysis method. The separation of experimental pre-edge peaks A and B
is larger by $\sim 0.8$~eV in comparison with the distance obtained
theoretically. We explain it by the fact that the theoretical width of
the energy band gap of the GaAs host material is smaller than the
experimental width.


Experimental data obtained for the non-etched annealed sample show an
increase in the intensity ratio between the highest peak C and lowest
peaks A, B in comparison with the etched sample
(Fig.~\ref{exp-anneal}). 
The shoulder above the absorption edge is observed only for high $\alpha$, 
the main peak C becomes sharper. Such phenomenon can be explained
by combined influence of both the content changes of Mn defects in
(Ga,Mn)As and the presence of the oxide layer on the sample
surface. To illustrate our assumption, we add an additional component,
the Mn K-edge XANES spectrum of MnO oxide, to the linear 
combination of three Mn model spectra in (Ga,Mn)As, 
and plotted the fit together with the
experimental curve, as depicted in Fig.~\ref{exp-anneal}. 
The main peak C in the fit becomes sharper, the shoulder smooths,  
and the intensity ratio between the main
and pre-edge structures reduces 
at the expense of the Mn oxide component


\section{Conclusion \label{Concl}}
We have performed an extensive LAPW numerical study of 
K-edge XANES spectra of the substitutional and 
two tetrahedral interstitial Mn sites in (Ga,Mn)As supercells. 
We have found out that the
effect of the core hole is essential for Mn$_{Ga}^{sub}$. 
We studied the whole range of absorption spectra, the highest
intensity Mn absorption lines in the main part of XANES above the
absorption edge and weak lines in he pre-edge part of XANES where the
sharp distinction between the spectra with substitutional and
interstitial Mn defects has been observed.
The difference between simulated spectra is determined mainly by 
the second nearest neighbor ligands.
Two peaks appear in the pre-edge region for any interstitial defect,
whereas a smaller single peak is obtained for a substitutional Mn
impurity. All pre-edge peaks are small in intensity in comparison with
the corresponding highest peak C in the main part. The combination of
theoretical spectra for various contents of Mn atoms in both
substitutional and interstitial positions gives a reasonable
qualitative fit of experimental data collected on the annealed 
and etched sample.
Summarizing the experimental and theoretical evidence, we have revealed
that even after annealing and etching of our sample some Mn atoms
still reside in interstitial sites, as spectra with two small peaks
are always observed experimentally.
                  

\section{Acknowledgements}
The assistance of Dr. S.\ Pascarelli from the beamline BM28 at ESRF
Grenoble is highly appreciated.
This work has been supported by the 
Ministry of Education of the Czech Republic (Center for
Fundamental Research LC510 and the research program MSM0021620834), 
and Academy of Sciences of the Czech Republic project KAN400100652.

\end{document}